# Pulsed laser deposition of rocksalt magnetic binary oxides

Alireza Kashir[1*], Hyeon-Woo Jeong[1], Gil-Ho Lee[1], Pavlo Mikheenko[2], and Yoon Hee Jeong[1*]

[1]Department of Physics, Pohang University of Science and Technology (POSTECH), Pohang, 37673, Republic of Korea

[2]Department of Physics, University of Oslo, P.O. Box 1048 Blindern, 0316 Oslo, Norway

**Corresponding Authors:**

*yhj@postech.ac.kr
*kashir@postech.ac.kr

## Abstract

Magnetic binary oxides with the rocksalt structure constitute an important class of materials for potential applications as electronic or electrochemical devices. Moreover, they often become a theoretical playground, due to the simple electronic and crystal structures, in the quest for novel phenomena. For these possibilities to be realized, a necessary prerequisite would be to grow atomically ordered and controllably-strained binary oxides on proper substrates. Here we systematically explore the use of pulsed laser deposition technique (PLD) to grow three basic oxides that have rocksalt structure but different chemical stability in the ambient atmosphere: NiO (stable), MnO (metastable) and EuO (unstable). By tuning laser fluence $F_L$, an epitaxial single-phase NiO thin-film growth can be achieved in a wide range of growth temperatures $10 \leq T_G \leq 750$ °C. At the lowest $T_G$, the out-of-plane strain raises to 1.5%, which is five times higher than in NiO film grown at 750 °C. MnO thin films that had long-range order were successfully deposited on the MgO substrates after appropriate tuning of deposition parameters. The growth of MnO phase was strongly influenced by $F_L$ and the $T_G$. EuO films with satisfactory quality were deposited by PLD after oxygen availability had been minimized. Synthesis of EuO thin films at rather low $T_G$ = 350 °C prevented thermally-driven lattice relaxation and allowed growth of strained films. Overall, PLD was a quick and reliable method to grow binary oxides with rocksalt structure in high quality that can satisfy requirements for applications and for basic research.

# 1. Introduction

Magnetic binary oxides with cubic rocksalt structure and the molecular formula MO, where M is a magnetic metal, like MnO, NiO, CoO, FeO, EuO, etc. form a group of insulating materials with antiferromagnetic alignment of atomic magnetic moments achieved by the superexchange interaction [1-3], except EuO, which is a ferromagnet [4]. These oxides are the best candidates for studying electronic behavior and exploring new phenomena that emerge in strongly-correlated electronic systems. For decades, they attracted a huge attention in solid-state electronics. The simplicity of their crystal, atomic and electronic structure enable theorists to mathematically calculate and analyze their electronic behavior under diverse conditions. Moreover, these materials are the best in which to experimentally investigate the accuracy of the new mechanisms proposed for the electronic behavior in solid-state physics; e.g., spin-phonon coupling [5-8], magnetic ordering [9] and insulator-metal transition [10-11]. The role of this family of oxides in the development of 20$^{th}$-century physics is such remarkable that it is not an overstatement to claim that the physics owe them greatly for its dazzling progress.

Application of strain can create a range of systems with a continuous change of new physical properties in materials [12-14]. Examples include insulator-metal transition, transformation of non-ferroelectric materials to ferroelectrics, and creation of multiferroic materials by increasing spin-phonon coupling [15-20]. Application of strain and induction of point defects can induce appearance of interesting features in magnetic binary oxides that have the rocksalt structure. A highly-compressed NiO single crystal passes through an insulator-metal Mott transition [10, 20], which is expressed as a strong nonlinear decrease in resistance by about three orders of magnitude. Application of appropriate strain can induce ferroelectricity in a binary compound, specifically in EuO [15, 17, 21] Magnetically-induced electric polarization may occur in a highly strained magnetic binary oxide (e. g. MnO) that includes a superexchange chain of atoms [19].

Thin-film deposition is a technique that can be used to control strain in materials [22]. The Curie temperature $T_C$ of EuO thin film is strongly dependent on the strain and on film thickness $t_F$ [23]. Dislocations in NiO thin films locally change the magnetic ordering from antiferromagnetic to ferromagnetic [24]. Such 'ferromagnetic' dislocations originate from a local non-stoichiometry at in dislocation cores where Ni is deficient. A slight deviation from the stoichiometry also changes the conduction behavior in NiO thin films. Oxygen-deficient EuO shows metallic behavior with a substantial increase of the $T_C$ to 120 K [25]. The deposition of high-quality magnetic binary oxides, and the ability to control the level of strain and density of defects in their structure, are challenging tasks in materials science and solid-state physics.

Pulsed-laser deposition technique (PLD) has many advantages for growing thin films of high-quality oxides [26-31]. PLD can preserve the stoichiometry of the film composition, can grow complicated oxides and deposition of ceramic thin films at relatively low growth temperatures $T_G$, and can provide opportunities to fabricate many different systems and to design unique electronic properties in the materials. However, this method applies high-energy pulses which may cause undesirable defects in the structure, and thereby change the electronic behavior of the films. PLD has been used to deposit various binary compounds [32-47].

Epitaxial NiO films can be obtained by PLD technique in a wide range of $T_G$ (even at room temperature) [33-34] and in a broad range of oxygen pressures [38]. Hetero-epitaxial NiO film on MgO substrate was successfully grown using PLD at room temperature [33]. Oxygen pressure $P_{O2}$ inside the chamber affects the film quality and lattice parameters of NiO by affecting the development of Ni deficiencies [38]. To grow epitaxial NiO film an optimum $P_{O2}$ should be inserted into the chamber [33, 34]. In a given $T_G$ the width of rocking curve around NiO diffraction peak was substantially changed as a function of $P_{O2}$ which indicates a considerable change in crystalline quality of NiO films [34, 40]. The $T_G$ appeared to be a crucial parameter on obtaining high quality NiO film [38, 41]. Increasing the temperature greatly decreases the surface roughness of deposited films [32]. It is proven that there should be an optimum temperature to grow NiO with highest quality on $Al_2O_3$ substrate [38]. Increasing the laser fluence $F_L$ strongly affects lattice parameters of NiO film grown on $Al_2O_3$ [0001] substrates by increasing the Ni-Ni bond length [39]. By referring to previous works and by tuning deposition parameters we have tried to obtain NiO films with atomically long-range ordering and a perfect smooth surface, and to study the effect of thickness and $T_G$ on the structural evolution and strain accommodation in the film structure.

MnO seems to be more complicated to grow than NiO, because Mn-O phase diagram has four phases: MnO, $Mn_3O_4$, $Mn_2O_3$ and $MnO_2$ [49]. This diversity restricts the options to grow high-quality single-phase MnO. $P_{O2}$, $T_G$ and $F_L$ should be set carefully to achieve single-phase MnO film. Although Single-phase MnO is achievable by PLD [42-45], but the effects of different PLD parameters on phase characteristics and structural morphology have not been studied yet. Instead, more focus was on how to achieve single phase MnO preventing the growth of other phases in Mn-O system. Increasing $P_{O2}$ inside the PLD chamber considerably changes the composition of grown films [45]. Here, we first try to find the best condition to avoid growing any other phase except MnO, and then we tune the parameters to achieve thin film that has atomically long-range order.

EuO is unstable in ambient conditions [50]. Molecular beam epitaxy (MBE) is the conventional method to grow EuO film. Growing strained single-phase EuO with reasonable structural quality by PLD is one of the most challenging topics in the film deposition. PLD-grown EuO has been already achieved by several research groups [25, 46-47] but due to the severe restricted condition the effects of the most important parameters, i.e. $P_{O2}$ and $T_G$ have not been studied in detail. This lack motivates us to perform a comprehensive study on the growth behavior of EuO on different substrates by PLD system.

Each of these oxides may represent a group of oxides that have the same stability. A valuable information about how to induce strain and achieve long-range atomic order in the films, will be revealed.

## 2. Experiment
### 2.1. Substrate preparation

As-received substrates, $SrTiO_3$ (001) (STO), MgO (001) (MgO), $LaAlO_3$ (001) (LAO) and $YAlO_3$ (110) (YAO) (all from Crystech) were cleaned ultrasonically in acetone for 10 min then in methanol for 10 min to remove contaminants from the top surface. STO and LAO substrates, were ultrasonicated for 5 min in deionized water then etched for 30 s in buffered hydrogen fluoride ($NH_4F$: HF=7:1, pH~ 4.5) then for 30 s in buffered HCl (pH~ 4.5). The substrates were dried under flow of high-purity $N_2$ gas, then STO, LAO and YAO were annealed at 1000 °C for 2.5 h in air; MgO was annealed at 1150 °C for 3 h.

### 2.2. Film Deposition
#### 2.2.1. Nickel Monoxide

NiO film was grown by using a NiO polycrystalline ceramic pellet (99.99% purity) as a target. An Nd-YAG laser (266 nm) with $F_L$ = 2 J/cm$^2$ was used for film deposition. Films were grown on $TiO_2$–terminated STO (001) substrates at substrate temperatures $10 \leq T_G \leq 750$ °C. To compensate for the expected oxygen deficiency in the deposited materials, the chamber was filled with high-purity (99.999%) $O_2$ gas at 2 Pa. After deposition, the films were cooled to RT at 12.5 °C/min.

#### 2.2.2. Manganese Monoxide

MnO film was grown by using a polycrystalline MnO ceramic pellet (99.99% purity) as a target. A KrF laser (248 nm) was used for the film growth. All samples were grown in vacuum of 13×10$^{-5}$ Pa. Various combinations of $T_G$ and $F_L$ were tested to achieve the best conditions for growing MnO film without inclusions of other phases. The Mn-O system includes four members (MnO, $Mn_3O_4$, $Mn_2O_3$, $MnO_2$) [49] so tuning the deposition parameters to grow pure rocksalt binary oxide MnO proved to be a challenging task. After deposition, the films were cooled to RT at 10 °C/min.

#### 2.2.3. Europium Monoxide

Pulsed laser deposition of EuO is more complicated than growth of NiO and MnO, because both Eu and EuO are extremely sensitive to oxygen [50]. Therefore, the chosen deposition target had reduced oxygen content. $Eu_2O_3$ polycrystalline ceramic target (99.99% purity), which was later replaced with a Eu target of 99.9% purity. A KrF excimer laser with the wavelength of 248 nm was also used for this deposition. $T_G$, substrate material and $P_{O2}$ were the most important parameters during deposition of EuO. Growth of EuO requires ultra-high vacuum; we used a base pressure of 13.3×10$^{-7}$ Pa. To take advantage of nonequilibrium processes, $F_L$ = 2.5 J/cm$^2$ was used to deposit the samples. To prevent degradation of grown films by air during the measurements, they were capped with a ~ 2-nm MgO layer right after deposition, at the same conditions of growth. After deposition, the films were cooled to RT at 10 °C/min.

In this work for all depositions, the optimum target-substrates distance $D_{TS}$ was chosen 50 mm, and the repetition rate of the laser pulses was 10 Hz.

## 2.3. Characterization
### 2.3.1. Surface morphology

The surface morphology of the substrates and the deposited samples was investigated using Atomic Force Microscopy (AFM) in dynamic non-contact mode on a Park system XE150 scanning probe microscope.

### 2.3.2. Phase analysis and crystal structure

The crystal structure of the films was characterized by X-ray diffraction (XRD) using CuK$_\alpha$ radiation from a source operating at 40 kV with current of 200 mA. This study was followed by an open detector ω scan around the detected Bragg peaks to investigate the configuration of the crystalline lattice in detail. An X-ray reflectometer (XRR) was used to measure $t_F$ and the quality of the surface and of the film/substrate interface. To estimate $t_F$, we considered the first five oscillations in the XRR pattern and used equation (1) to obtain four different values.

$$t_F \sim \frac{\lambda}{2} \frac{1}{\theta_{m+1}-\theta_m}, \qquad (1)$$

Where λ is wavelength of X-ray, $\theta_{m+1}$ and $\theta_m$ are the position of *(m+1)-th* and *m-th* interference maximums, respectively. Then $t_F$ was obtained by taking the average of these values and then rounding it to nanometer scale. When Pendellosung oscillations were observed around the detected Bragg peaks, we used these oscillations to determine $t_F$ as

$$t_F = \frac{\lambda}{2(\theta_{m+1}-\theta_m)cos\theta_B}, \qquad (2)$$

Where $\theta_B$ is the detected Bragg peak.

## 3. Results and Discussion
### 3.1. Substrates

AFM revealed steps of atom-scale height, and flat terrace structure (Fig. 1) on the surface of all treated substrates [51-54]. YAO showed an atomically smooth surface, as proven by a line scan through its surface. MgO substrate with a step-terrace surface can be obtained after vacuum annealing at 950 °C [55], but this process may cause formation of point defects, which may change the electronic structure on the substrate surface. In our treatment, the annealing environment was changed to the air, and to avoid the hindering effect of atmospheric pressure on the surface diffusion, the annealing temperature was increased; several experiments determined that 1150 °C is an optimal annealing temperature for surface treatment of MgO (001).

### 3.2. Nickel monoxide

XRD study on NiO films grown at 750 °C under pressure of 2 Pa oxygen shows a single peak of NiO (002) from $2\theta$ = 20 to 80° which reveals that the films' crystallographic planes grow to match the orientation of the substrate. The scan did not show additional peaks that indicate the presence of any other phases, like $Ni_2O_3$ or metallic Ni. For clear visualization, we show a limited range of $2\theta$ in which the evolution of NiO (002) peak versus thickness is clearly observable (Fig. 2a).

As $t_F$ decreased, the NiO (002) peak shifted to the left; this change is a result of out-of-plane elongation of NiO lattice structure. This trend shows that the deposited films may be under in-plane compressive strain due to the lattice mismatch between NiO and STO. Repulsion of similar charges induced by point defects (like Ni vacancies) could also be a reason for initial lattice expansion in PLD-grown films. The highest achieved out-of-plane strain was 0.5% at $t_F$ = 5 nm, and relaxation was nearly total at $t_F$ =200 nm [Fig. 2a, inset]. The relaxation processes start in the early stages of growth, so that the films cannot reach the theoretically-predicted strain of 6.5%. Indeed, NiO thin films grown on the $SrTiO_3$ substrates undergo fast relaxation of strain at the 750 °C temperature that was used for growth here, possibly because the adsorbate-adsorbate interaction is stronger than the adsorbate-substrate interaction [56]. The $t_F$ itself seems to be not a crucial parameter for the relaxation process in NiO films grown at 750 °C. However, $T_G$ is likely to influence the fast relaxation process. The hot substrate provides thermal energy for the incoming ablated particles; this energy is sufficient to overcome the diffusion barriers at the surface, and therefore accelerates the surface diffusion and ultimately releases the mismatch strain between the NiO film and the $SrTiO_3$ substrate. However, the critical $t_F$ for strain relaxation ($h_c$) decreases as theoretical lattice mismatch $\alpha_{th}$ increases, so the large $\alpha_{th}$ = 6.5% between NiO and STO also accelerates the relaxation process.

Increase in $F_L$ causes increase in out-of plane lattice parameters in NiO thin films grown on $Al_2O_3$ (0001) [39]. In our investigation, under a constant $F_L$ the strain accommodation in the film decreased as $t_F$ increased, and a NiO film that had $t_F$ = 200 nm was totally relaxed.

The ω rocking scan around NiO (002) peak yields additional information on the structural features of the grown films. The shapes of the curves changed as $t_F$ changed (Fig. 2b).

Films that had $t_F$ = 2.5 nm had only one sharp peak that has width < 0.05° which is instrumental limited. Films that had $t_F$ > 40 nm show a simple broad peak. The two components imply the existence of two different structural correlation lengths throughout the films [57-62]. The sharp peak reveals an atomically long range-ordered area. This feature has been never seen in previous works on NiO films [32-41]. The emergence of the broad component shows the formation of sub-grains (mosaic structure); this process breaks the long-range atomic arrangement. The rocking curves of films that had 2.5 nm $\leq t_F \leq$ 40 nm showed a combination of these components. However, as $t_F$ decreased, the width of broad peak increased (Fig. 2b, inset) and the sharp peak appeared and intensified. The ratio of the broad to sharp components' intensities was negligible for $t_F$ < 5 nm. These changes indicate the spread of the long-range ordered phase and reduction in mosaic sizes. Threading dislocations become increasingly common as $t_F$ increases [60]; they can disrupt a long-range ordered area by forming subgrain boundaries (forming mosaic structure), so this process may cause the development of the broad peak. The increase in strain relaxation as $t_F$ increases might be a result of an emergence of dislocations at the interface between the film and the substrate as well.

The surface morphology of the film changed as $t_F$ increased from 15 nm to 200 nm, from a perfect steps-and-terraces (Fig. 3a) structure to an atomically smooth-surface without clear steps (Fig. 3d) then to an island-like structure (Fig. 3e). During initial stages of growth, the deposited materials seemed to follow the atomic structure of the substrate, and that gradually the clarity of the steps faded. The gradual change of growth surface from a perfect steps-and-terraces structure of STO to a defective NiO crystal might be the reason behind different growth behaviors of the films. Defects influence this behaviour by providing nucleation sites as the energically prefered place for growth of the films.

For further investigation, NiO thin films were deposited at a wide range of $T_G$, while the other parameters were held the same as for the films described above.

XRD of a film grown at 10 °C showed a single NiO (002) peak (Fig. 4a). This feature helps to study the strain relaxation behavior as a function of $T_G$. Calculation of the out-of-plane lattice constant from the position of the Bragg peak determined that the out-of-plane strain in an NiO film that has $t_F$ ~ 15 nm is five times higher when it was grown at 10 °C than when it was grown at 750 °C (Fig. 4a, inset). This observation is consistent with the hypothesis that the fast strain relaxation is due to thermal energy that is provided by the substrate during the growth. Therefore, to obtain a strained NiO thin film, $T_G$ should be kept low.

Reduction of $T_G$ strongly altered the general peak of the ω rocking curve around the NiO (002) diffraction peak (Fig. 4b). The sharp peak was clear in the film grown at $T_G$ = 750 °C, but was totally absent at $T_G$ = 600 °C. This change indicates that the atomically long-range ordered area disappeared as $T_G$ was reduced, and that the mosaic blocks spread throughout the film. The width of the broad peak increased as $T_G$ decreased (Fig. 4b, inset); this change is an sign of decrease in crystallite sizes in the mosaic structure, and therefore, of degradation in film quality. The decrease in thermal energy with reduction in $T_G$ would decelerate the surface diffusion of particles supplied to the substrate; this change might be responsible for the resulting smaller crystallite sizes in films grown at low $T_G$. AFM image showed an atomically-flat steps-and-terraces surface structure in an NiO thin film grown at 10 °C (Fig. 5a). XRR patterns from films with $t_F$ ~ 15 nm grown at different $T_G$ confirm that all have high-quality surfaces and interfaces with substrates (Fig. 5b).

### 3.3. Manganese Monoxide

To form pure MnO, $T_G$ must be high and $P_{O2}$ must be low to avoid forming the other phases [49]. Normally these conditions is not achievable in the PLD chamber, but by taking advantage of the non-equilibrium processes, the fabrication can be achieved in certain conditions.

To deposit MnO, we started by setting $T_G$ = 750 °C and placing the substrate 50 mm from the target; then tried to achieve deposition of MnO single phase by testing combinations of $P_{O2}$ and $F_L$. Both parameters at the given deposition conditions strongly influence the quality and composition of the films.

XRD data (Fig. 6) indicate that, to grow a pure MnO phase in vacuum of 13×10$^{-5}$ Pa, the $F_L$ should be < 1.5 J/cm$^2$. When $F_L$ was increased to 3 J/cm$^2$, MnO$_x$ phases with $x > 1$ emerged, possibly because of generation of thermally-activated fragments that had high reactivity with oxygen. As a result, the Mn$^{3+}$ valence sites appeared in the film. At $F_L$ $\geq$ 4 J/cm$^2$, the XRD scan showed no film peak; i.e., the layer was amorphous or no material was deposited. Re-sputtering of the ablated species due to their high kinetic energy can be a reason for this result [63], because of a scarcity of background gas to reduce the kinetic energy of particles when they are landing on the substrate surface. Our investigation shows that regardless of the substrate-target distance, the film composition changes from pure MnO to Mn$_3$O$_4$ as $F_L$ increases.

$T_G$ affects the quality and purity of MnO thin films in two ways. First, increase in $T_G$ accelerates surface diffusion, so well-crystallized film can form. Second, $T_G$ affects the proportions of MnO$_x$ ($x > 1$) in the film [47]. XRD patterns (Fig. 7) were obtained for films grown at 750, 650, 600 or 550 °C. As $T_G$ was decreased to 550 °C, the Mn$_3$O$_4$ phase emerged, as predicted by the Mn-O phase diagram. At the same time, the position of the MnO (002) peak shifted to the left; this change is a result of out-of-plane lattice expansion. The lattice mismatch of the film and substrate, and the point charges created by the defects, might be responsible for this expansion. This result demonstrates that to avoid the formation of any phases other than MnO in the deposited film, $T_G$ should be > 650 °C. At $T_G$ > 650 °C, only MnO phase formed; it grew in the 002 direction, which follows the crystallographic orientation of the substrate.

MnO films had poorer quality when grown at 750 °C than at 650 °C (Fig. 8). ω rocking around the (002) Bragg peak for ~ 25-nm MnO film grown at 750 °C showed a single broad peak (FWHM=1.46°), which is a signature of mosaic structure. This structure is a typical feature of films deposited in a vacuum [42], because in this condition no gas exists to exert decelerating force that reduces the kinetic energy of ablated species. Subsequently, the highly energetic particles create defects in the grown films and break their atomic order. Large theoretical lattice mismatch between MnO and MgO also causes forming mosaic area in deposited film. Reduction of temperature to 650 °C was accompanied by the emergence of a narrower peak (FWHM=0.16), which indicates the formation of an atomically longer range ordered area. This phenomenon is completely against expectations, because thermal energy usually drives the system to increased atomic order. High $T_G$ might increase the sensitivity of the deposited material to the impact of subsequently-arriving particles during film growth. This process increases the defective content in the structure (Fig. 8c, d). The surface structure of the film grown at 750 °C is granular with the average grain size of about 250 nm, whereas a smooth surface appears for the film deposited at 650 °C. This analysis of crystal quality and phase purity of MnO, suggests the existence of an optimum $T_G$ for deposition.

### 3.4. Europium Monoxide

Elemental europium is so sensitive to oxygen that even in a fraction of a minute in ambient atmosphere, a polished surface oxidizes to $Eu_2O_3$. EuO is not stable under typical PLD conditions [50], so growth of EuO is a very challenging task. To establish a condition for deposition of pure single-phase EuO, this study focused on the most-effective parameters, especially $T_G$ and the amount of oxygen available in the PLD chamber. Usually, the PLD process is affected by oxygen in four ways: it is contained in the target material, it is present as background gas, it is a residual element in the vacuum chamber and it is a component of the substrate. In the deposition, $T_G$ was set to 350 °C; at this temperature, EuO is stable in a very narrow range of $P_{O2}$, which is difficult to achieve in PLD chamber. However, the non-equilibrium processes in the PLD can help in growing EuO even at relatively high $P_{O2}$.

First, we tried to grow EuO thin film at different conditions using a $Eu_2O_3$ ceramic target, but due to the highly-oxidized target, EuO composition could not be achieved even in high vacuum, pressure ~$10^{-8}$ Pa. XRD scans detected only $Eu_3O_4$ and $Eu_2O_3$.

To remove the major source of oxygen, we replaced the target with a high-purity Eu metal disk and grew samples under different conditions on three different substrates: LAO (001), YAO (110) and MgO (001). XRD patterns of films grown in vacuum of 6.7×$10^{-8}$ Pa at 350 °C showed single phase of EuO grown on LAO and YAO (Fig. 9), but in films grown on MgO substrate, the EuO (002) peak was not strong. This result is consistent with the hypothesis that ablated Eu particles form EuO by combining with oxygen from the substrate [47]. The surface structure of the substrate (lattice spacing and atomic arrangement) is also important for growing high-quality thin films. The Lattice mismatch was < 4% on YAO (110) and LAO (001), but 18% on MgO (001) in both x and y directions. The XRD pattern of EuO film on YAO substrate showed Pendellosung oscillations, which were absent in the XRD pattern of film grown on LAO. This difference indicates that the film surface and interface were of higher quality between EuO and YAO than between EuO and LAO, possibly as a result of the lower lattice mismatch in EuO/YAO (~ 2%) than in EuO/LAO (~ 4%). Using these oscillations by applying Eq. 2 the film thickness was calculated to be ~ 10 nm. The ω rocking scan around EuO 002 Bragg peak indicates a considerable effect of the type of substrate on crystalline features of deposited films. Figure 10 shows a remarkable improvement in film quality when the substrate was changed from LAO to YAO. The width of curve decreased ~ 12 times (from 2.8° to 0.23°) which implies a considerable improvement in structural quality of the film.

When $T_G$ was reduced to 300 °C, the $Eu_2O_3$ phase emerged (Fig. 11). Our investigation shows that 350 °C is the most appropriate temperature to grow high quality EuO under the stated conditions.

Decrease in $t_F$ was accompanied by a right-shift of the EuO 002 peak in films grown on LAO substrate; this shift indicates ~ 3% out-of-plane EuO lattice compression, and suggests that the EuO film is under in-plane tensile strain due to the lattice mismatch with substrate (Fig. 12). The growth of high-quality single-phase EuO at relatively low $T_G$ enabled introduction of a high level of strain into the film structure, because of a relative lack of thermal energy to facilitate strain relaxation.

The purity of deposited films increased with increase in the time for which the target surface was cleaned by laser pulses before the growth process (Fig. 13). Oxidation of the Eu surface inside the chamber before film deposition may result in the presence of $Eu_2O_3$ phase in the deposited film. To grow pure EuO phase, the oxide layer must be removed from the target.

XRR scan of a film that had $t_F$ ~ 26-nm and had been grown at $T_G$ = 350 °C on YAO (110) substrate shows a good surface and interface quality (Fig. 14a). AFM showed that this film was homogeneous and had a smooth surface (Fig. 14b).

## Conclusions

Films of pure magnetic binary oxides (NiO, MnO, EuO) were deposited using pulsed laser deposition. Their purity, crystal structure and surface morphology were investigated using X-ray diffractometry and atomic force microscopy. By setting the growth parameters properly, PLD can deposit high-quality binary oxides even in high vacuum. The growth of epitaxial NiO thin films in a large range of temperatures enabled us to study the evolution of crystal structure and the strain relaxation as a function of $t_F$ and $T_G$. $T_G$ can greatly affect the strain relaxation in NiO thin films. To obtain pure MnO film, $T_G$ and $F_L$ appeared to be the crucial parameters; by tuning them, an atomically long-range ordered film was deposited. EuO is unstable and very sensitive to oxygen, this structure was grown in a typical PLD chamber condition by choosing appropriate target and substrate materials. Despite many instrumental limitations, a perfectly-crystallized strained EuO film was deposited by PLD.

## Acknowledgement

This work was partially supported by National Research Foundation (NRF) of Korea (2015R1D1A1A02062239 and 2016R1A5A1008184) funded by the Korean Government.

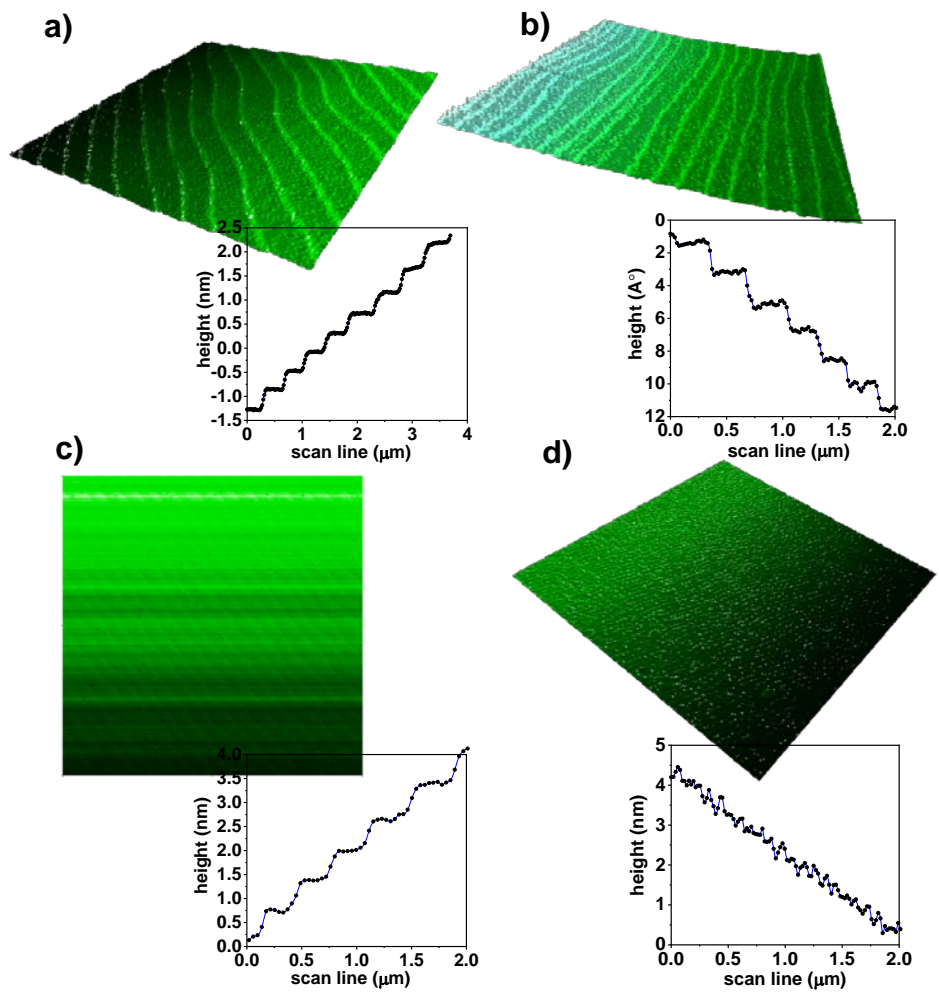

**Fig.1.** AFM (5 μm × 5 μm) topographic images of the treated substrates, (a) STO (001), (b) MgO (001), (c) LAO (001) and (d) YAO (110).

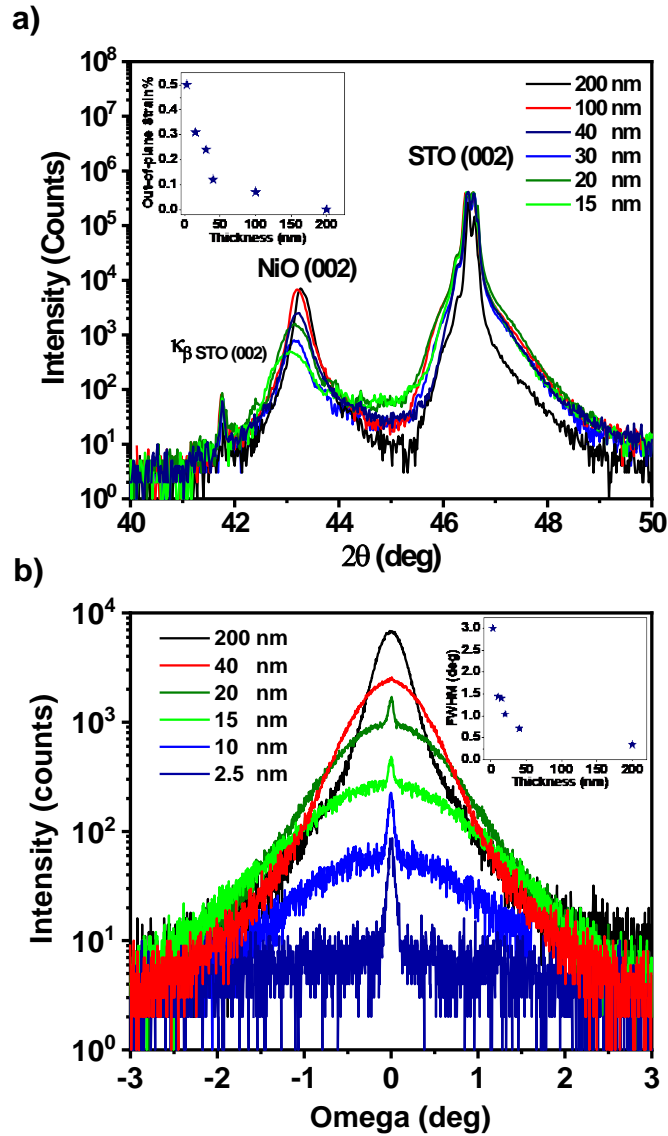

**Fig. 2.** (a) XRD patterns of NiO films with different thickness from 15 to 200 nm grown at 750 °C on STO (001) substrate, the inset shows the out-of-plane strain versus film thickness, (b) ω rocking curves around the NiO (002) diffraction peak, the inset shows the width of (002) NiO broad peak versus thickness.

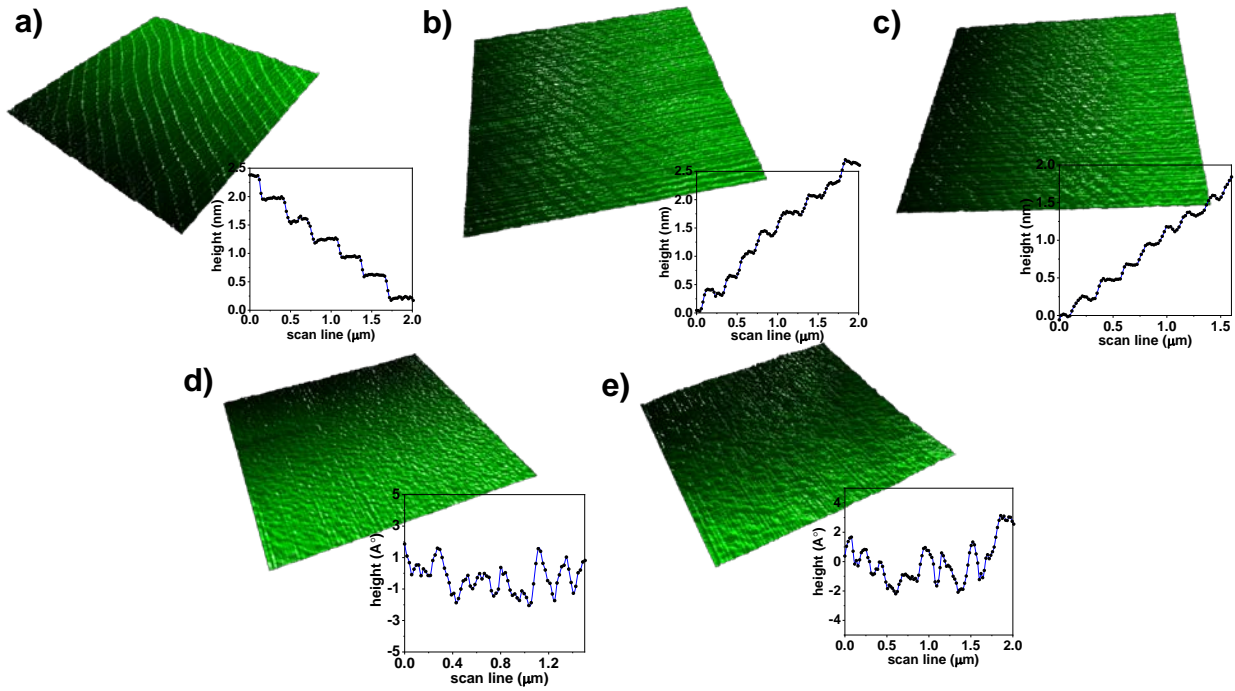

**Fig. 3.** AFM (5 μm × 5 μm) topographic images of NiO thin films with thickness of (a) 15 nm, (b) 20 nm, (c) 30 nm, (d) 100 nm and (e) 200 nm grown on STO (001) substrate at 750 °C.

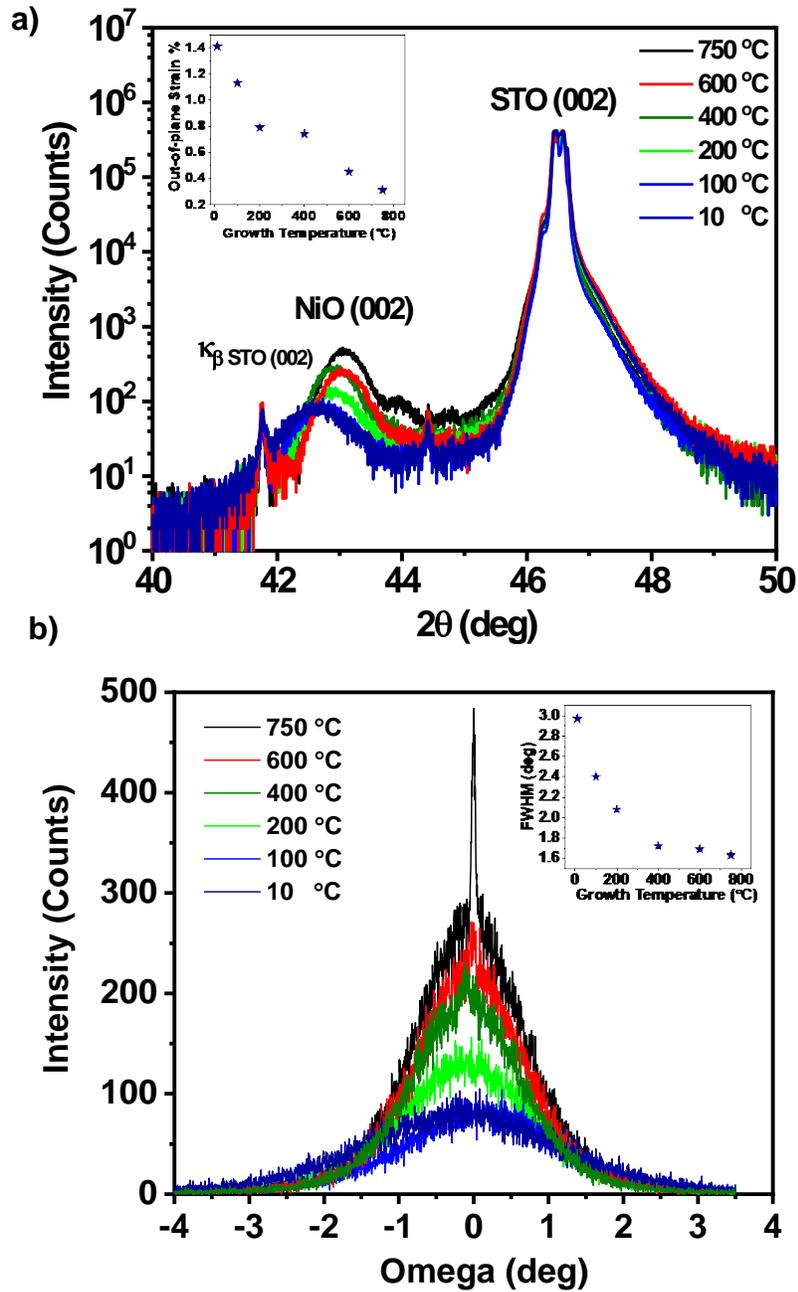

**Fig. 4.** (a) XRD patterns of the ~15-nm NiO films grown at different temperatures on the STO (001) substrate, the inset shows the out-of-plane strain derived from the NiO (002) peak position. (b) ω rocking curves around the NiO (002) diffraction peak, the inset shows the width of NiO (002) broad peak versus growth temperature.

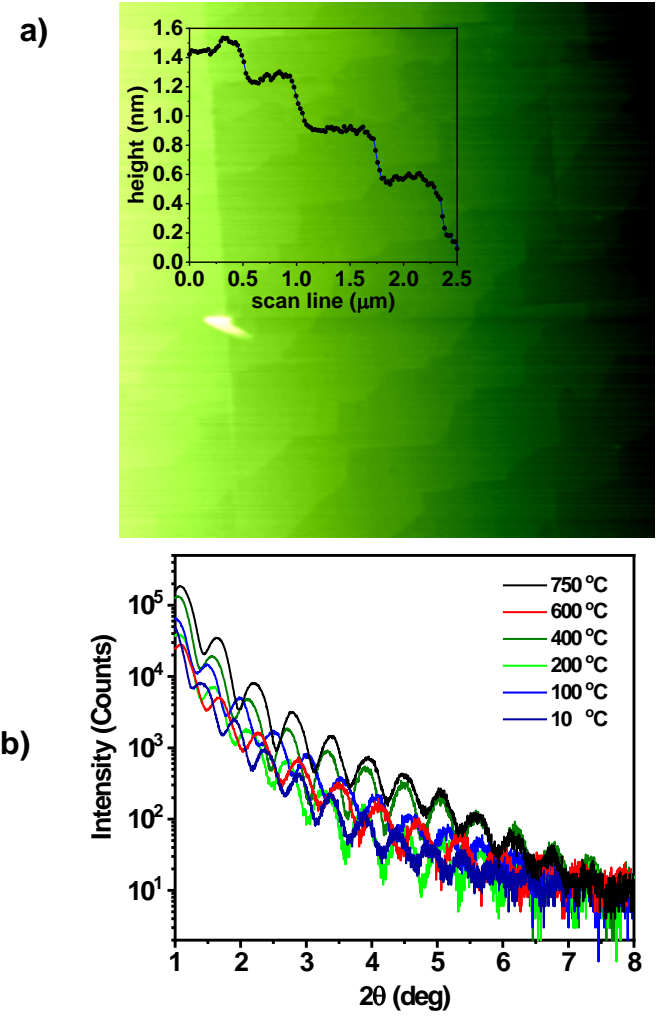

**Fig. 5.** AFM (5 μm×5 μm) images of ~15-nm NiO thin film grown on STO (001) substrates at 10 °C; (b) X-ray reflectivity of ~15-nm films grown at different temperatures.

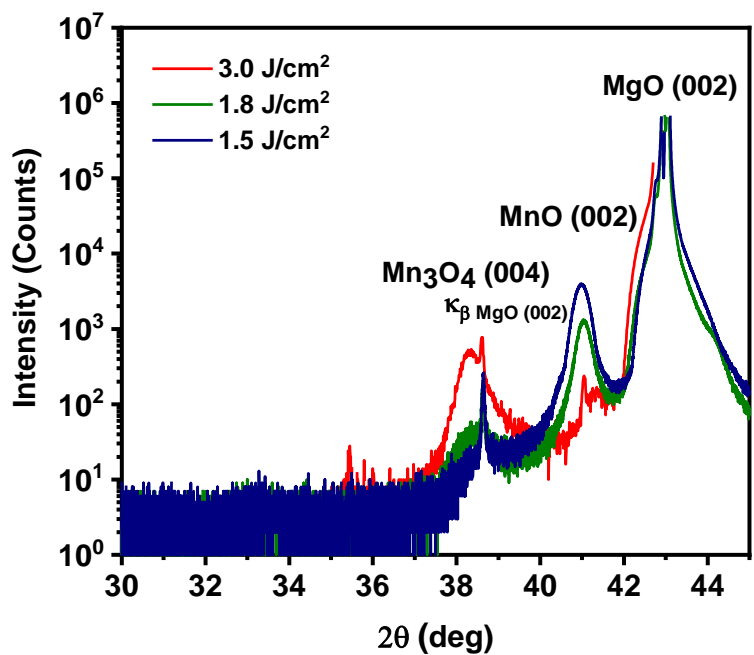

**Fig. 6.** XRD patterns of the MnO films grown at 750 °C with different laser energy fluences.

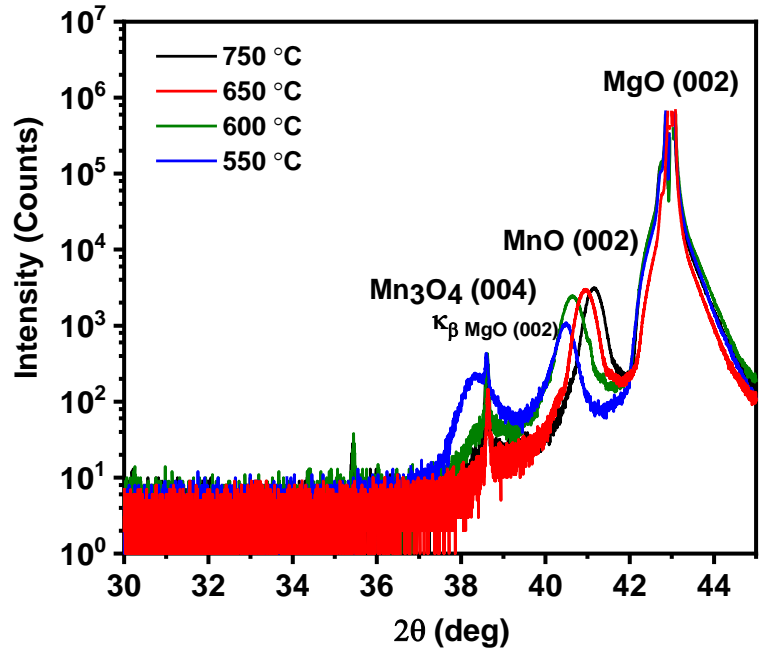

**Fig. 7.** XRD patterns of the MnO films grown at different substrate temperatures from 550 to 750 °C.

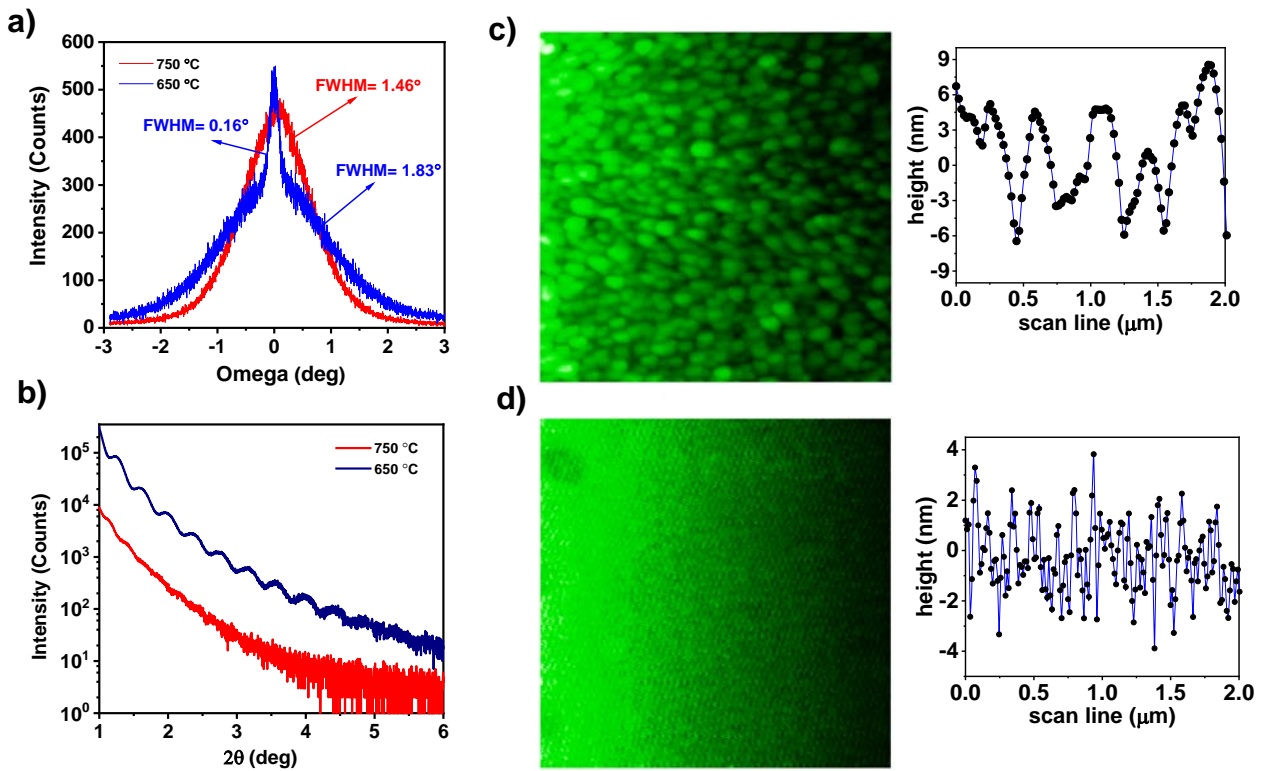

**Fig. 8.** (a) ω rocking scans around the MnO (002) Bragg peak of ~ 25-nm films, (b) X-ray reflectivity scan, AFM topographic images, for the ~ 25-nm MnO films grown on MgO (001) substrates at (c) 750 and (d) 650 °C, respectively.

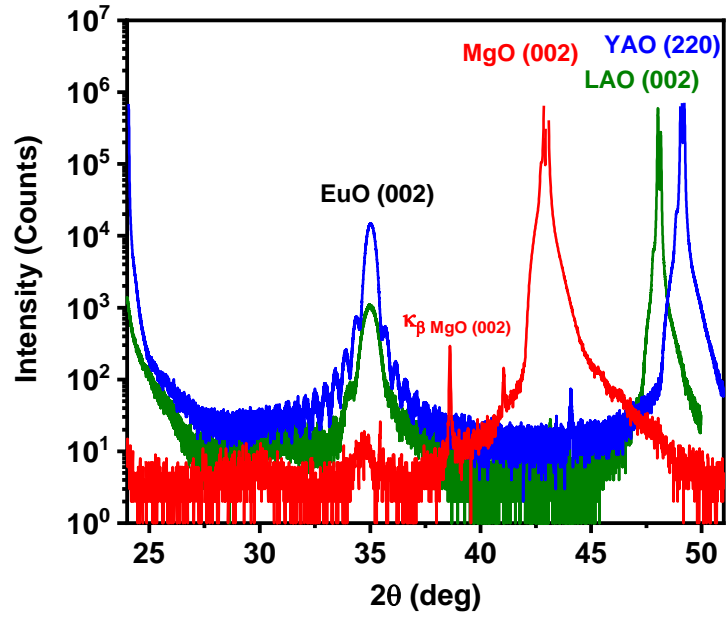

**Fig. 9.** XRD scans for EuO thin films grown on different substrates at 350 °C.

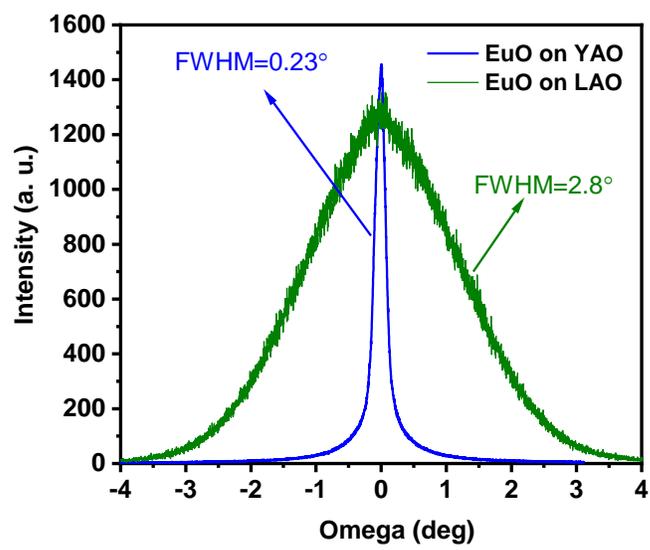

**Fig. 10.** ω rocking scans around the EuO (002) Bragg peak of ~ 25-nm films grown on different substrates at 350 °C.

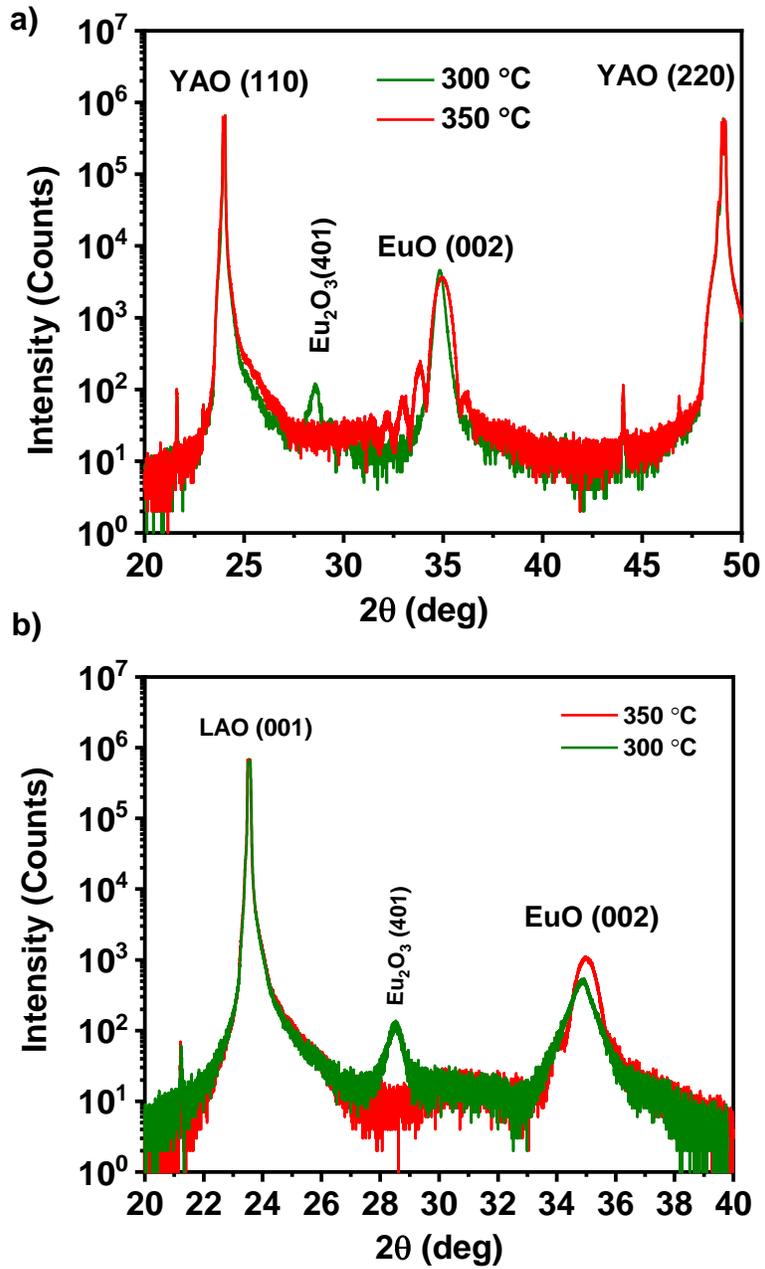

**Fig.11.** Effect of growth temperature on composition of deposited film on a) YAO (110), b) LAO (001) substrates.

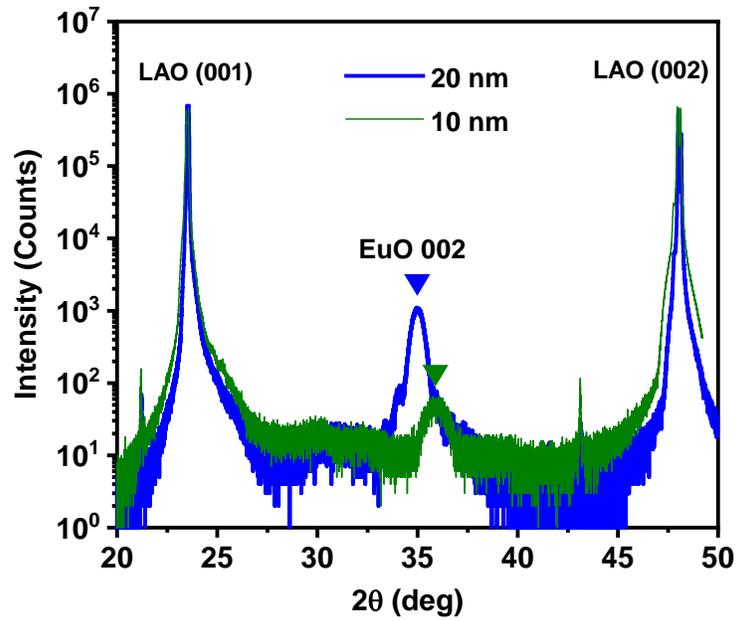

**Fig.12.** XRD patterns of EuO thin films with different thicknesses grown on LAO (001) substrate at 350 °C.

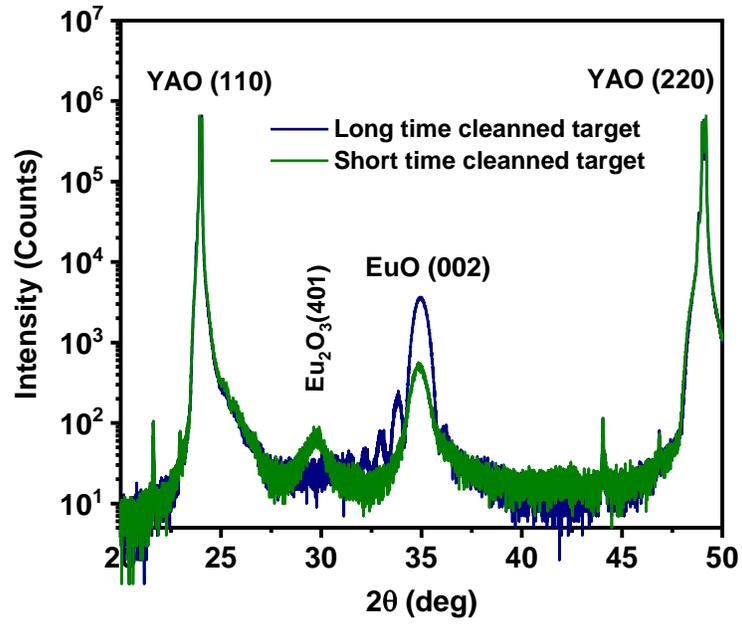

**Fig.13.** The effect of target cleaning time on the XRD patterns of the EuO films.

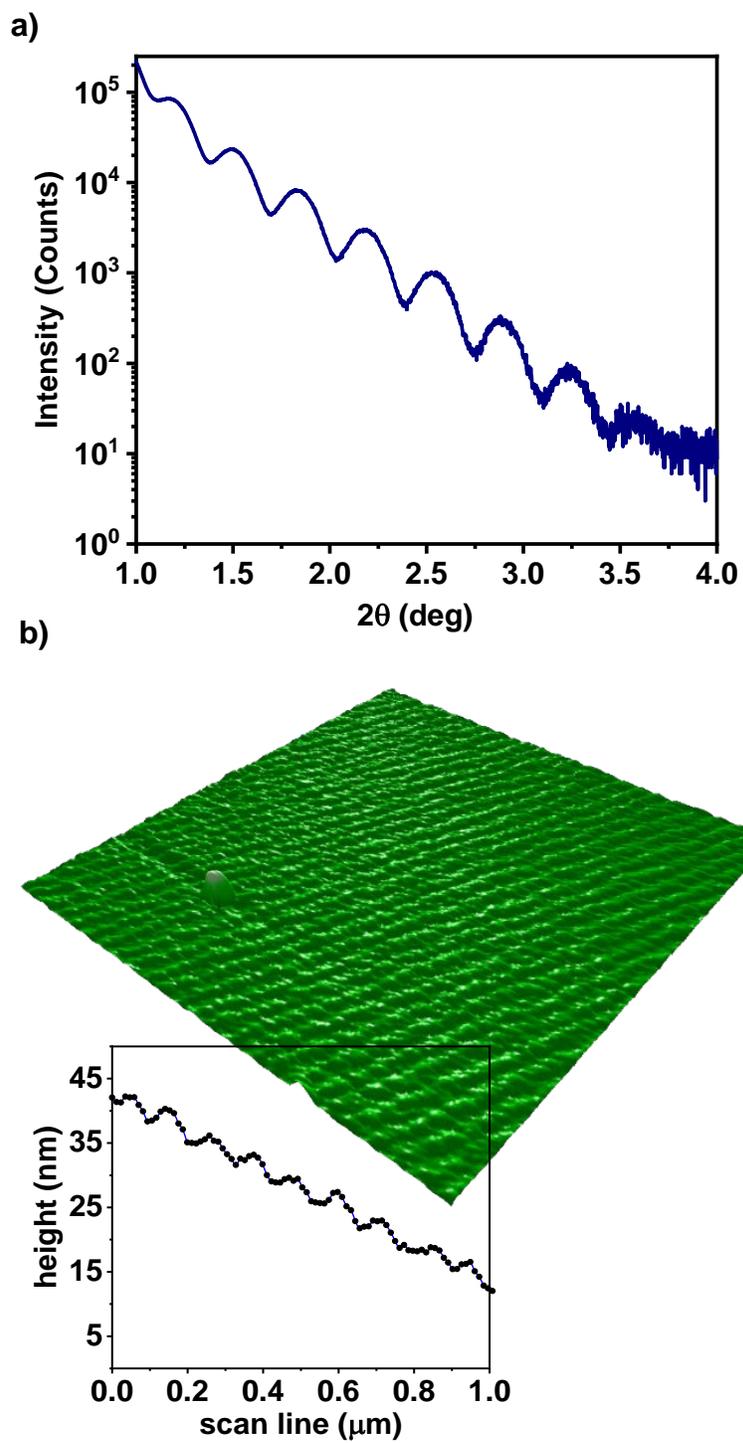

**Fig. 14.** (a) X-ray reflectivity scan for a ~26-nm EuO film deposited on YAO (110) at 350 °C (b) a 5 μm × 5 μm AFM topographic image of the film.